\documentclass[12pt]{article}
\usepackage{psfig}
\usepackage[cp1251]{inputenc} 

\begin{document}

\medskip

\centerline{\bf PHASELESS VLBI MAPPING OF COMPACT}
\centerline{\bf EXTRAGALACTIC RADIO SOURCES}

\bigskip

\centerline{A.T.Bajkova}

\bigskip

\centerline{\it Institute of Applied Astronomy, Russian Academy of
Sciences} \centerline{\it nab.Kutuzova 10, St.Petersburg,191187
Russia}

\bigskip

\noindent The problem of phaseless aperture synthesis is of
current interest in phase-unstable VLBI with a small number of
elements when either the use of closure phases is not possible (a
two-element interferometer)or their quality and number are not
enough for acceptable image reconstruction by standard adaptive
calibration methods.Therefore, we discuss the problem of unique
image reconstruction only from the spectrum magnitude of a source.
We suggest an efficient method for phaseless VLBI mapping of
compact extragalactic radio sources. This method is based on the
reconstruction of the spectrum magnitude for a source on the
entire UV-plane from the measured visibility magnitude on a
limited set of points and the reconstruction of the sought-for
image of the source by Fienup’s method from the spectrum magnitude
reconstructed at the first stage. We present the results of our
mapping of the extragalactic radio source 2200+420 using
astrometric and geodetic observations on a global VLBI array.
Particular attention is given to studying the capabilities of a
two-element interferometer in connection with the putting into
operation of a Russian-made radio interferometer based on Quasar
RT-32 radio telescopes.

\bigskip


~\vfill
\noindent E-mail: bajkova@gao.spb.ru

\newpage

\centerline{INTRODUCTION}

\vskip 7.5mm

The situation with uncertainty in measuring the phase due to
ionospheric and tropospheric inhomogeneity is typical of
phase-unstable interferometry. Phase errors, let alone the fact
that the phase is unknown, significantly restrict the signal and
image processing quality (Oppenheim and Lim 1981). Many studies
that are systematized in the monographs by Stark (1987) and
Thompson et al. (1986) are devoted to the influence of phase
errors.

In VLBI, adaptive calibration methods, including
hybrid and self-calibration methods that directly or
indirectly use closure phases (Cornwell and Fomalont 1999; Thompson et al. 1986),
are traditionally
used to solve the phase retrieval problem. The number
of equations for closure phases depends on the number of interferometer
elements $N$ and is $(N-1)(N-2)/2$ at each instant in time. The larger
the number
of interferometer elements and the more the visibility
measurements on each baseline, the higher the phase
retrieval accuracy. In contrast, at a small number of
elements and a small number of measurements, the
number of equations for closure phases may prove
to be insufficient for high-quality phase retrieval. For
a two-element ($N=2$) interferometer, there are no
equations for closure phases at all.

Phase retrieval errors can give rise to spurious features. Thus,
for compact extragalactic radio sources with a typical "core+jet"
structure, a spurious symmetric counterjet can appear on the map.
We encounter such a situation, for example, when mapping using
astrometric and geodetic observations on a global VLBI array
(IRIS, NEOS, etc.), as these are distinguished by relatively poor
UV coverage (Bajkova et al. 1997).

The necessity of mapping based on astrometric
data arises from the need to improve the coordinates
of reference sources by taking into account their
structure on milliarcsecond (mas) angular scales
(Bajkova 2002a). Clearly, because of their spectral
and temporal variability, the mapping of sources and
the astrometric reduction should be performed using
the same data. In addition, since the observations
are regular (each source has been observed once a
week for many years), they are also of considerable
interest in astrophysics, providing unique data for
investigating the structural evolution of extragalactic
sources on long time scales (Pyatunina et al. 1998).

Thus, apart from adaptive calibration methods,
phaseless mapping methods using only the visibility
magnitude can also be invoked to overcome the phase
uncertainty in VLBI, provided that the signal-to-noise is
high enough for the application of reconstruction algorithms
(see the section titled Reconstruction
Accuracy).

Interest in phaseless VLBI mapping is also being aroused by the
fact that the Institute of Applied Astronomy (Russian Academy of
Sciences)has recently put into operation a two-element
interferometer based on RT-32 radio telescopes of the Quasar VLBI
project at the Svetloe Observatory near St.-Petersburg and at the
Zelenchuk Observatory in the North Caucasus and that the first
preliminary mapping results have already been obtained (Bajkova
2002b).

In this paper, we present the results of our study of the
capabilities of a two-element interferometer designed to map
sources that can be represented as a set of compact components.
The results of this study can also be applied to
ground-based-spaceborne radio interferometers with the high orbit
of a space station (Finkelstein and Bajkova 1990).

The problem of aperture image synthesis without invoking phase
information was first considered and solved by Baldwin and Warner
(1976,1978). However, these authors suggested methods that were
applicable only in special cases where a source could be
represented as a limited number of point components and,
therefore, were not widely used in VLBI.

Here, our goal is to develop a more universal and efficient
phaseless image synthesis method that can be used for the VLBI
mapping of extragalactic radio sources, including those with a
"core+jet" structure, that, apart from compact components (core),
also contain extended components (jet); to study the capabilities
of a two-element interferometer; and to demonstrate the
potentialities of the suggested methods by mapping the well-known
source 2200+420 using astrometric and geodetic observations on a
global VLBI array as an example.

\vskip 7.5mm

\centerline{THE UNIQUENESS OF THE SOLUTION}

\vskip 7.5mm

In the most general formulation where constraints are imposed only
on the spectrum magnitude, the problem of reconstructing the
function has an infinite set of solutions. Indeed, any function
that has a given spectrum magnitude and an arbitrary spectral
phase satisfies these constraints, and, if at least one solution
is known, the other can be obtained by convolving this solution
with a function that has an arbitrary phase and a spectrum
magnitude equal to unity at all frequencies.

However, a significant narrowing of the set of solutions is
possible for certain constraints imposed on the function being
reconstructed in the spatial domain (Stark 1987). One of these is
the constraint imposed on the spatial extent of an object; i.e.,
the sought-for function must have a finite carrier. Another severe
constraint in the spatial domain is the requirement that the
solution be real and nonnegative. Below, in solving the phase
retrieval problem, we assume that the sought-for function
satisfies these constraints; i.e., it is real and nonnegative and
has a finite extent.

The finite extent of an object (the finiteness of the function)
ensures that the Fourier spectrum is analytic in accordance with
the Wiener-Paley theorem (Khurgin and Yakovlev 1971). As a result,
this spectrum can be reconstructed from the known part of it,
which is used to reconstruct images from the visibility function
measured on a limited set of points in the UV plane. If we
determine the class of equivalent functions to within a linear
shift and reversal of the argument (rotation through $180^o$),
then all of the functions that belong to this class have the same
spectrum magnitude. The solution of the phase retrieval problem is
assumed to be unique if it was determined to within the class of
equivalent functions.

In the case of one-dimensional functions, even these severe
constraints on finiteness and nonnegativity do not guarantee a
unique reconstruction from the spectrum magnitude. As the
dimensionality of the function increases ($n\ge 2$), a unique (to
within the class of equivalent functions) solution becomes
possible, except for the degenerate cases defined on the set of
measure zero (Bruck and Sodin 1979; Hayes 1982). This follows from
the qualitative difference between the properties of the $z$
-transformations of one-dimensional and multidimensional
sequences.

For a unique solution to exist,the $z$-transformation must be
irreducible, which is not achievable in principle in the
one-dimensional case and almost always holds in the
multidimensional case. Since we deal with two-dimensional images
in VLBI, we assume that the solution of the phase retrieval
problem exists and is unique. However, the existence of a unique
solution does not yet guarantee that the retrieval algorithms
converge. The papers by Gerchberg and Saxton (1972)and Fienup
(1978) are of greatest importance in developing the theory and
algorithms of solving the phase retrieval problem. Fienup’s
algorithm and its modifications aimed at speeding up the
convergence to the required solution (Fienup 1982) are most
efficient in terms of their applications. Based on numerous
simulations, we choose a hybrid input-output algorithm modified to
ensure convergence (by specifying the initial approximation) and
to prevent stagnation due to computational effects from the entire
set of various modifications of Fienup’s algorithm (Bajkova 1994,
1996).

\vskip 7.5mm

\centerline{A MODEL FOR THE STRUCTURE OF}

\centerline{COMPACT EXTRAGALACTIC RADIO SOURCES}

\vskip 7.5mm

Before turning to a description of the suggested phaseless mapping
method,let us consider simplified models for the structure of
compact extragalactic radio sources and their spectrum
magnitudes.The simplest approximation of a "core+jet" source
structure is a model that consists of two point components. This
model will be useful below when considering a two-element
interferometer. Let the brightest component be located at the
phase center of the map and be the core of the source (the
optically thick base of the jet) and the second component (the
optically thinner feature of the jet) have a lower brightness and
be located at an angular distance of $r$ mas with the position
angle $\theta$ reckoned from north to south through east
(clockwise). We will consider the relative brightnesses by
assuming the core brightness to be equal to unity. Let the
brightness of the second component be $A<1$.

The complex visibility function of such a source when its
individual components are represented as $\delta$- functions is
$$
V_1(u,v)=1+A\exp(2\pi i(u\xi_o+v\eta_o)),
$$
\noindent where $\xi_o=r\sin\theta, \eta_o=r\cos\theta$ are the
rectangular coordinates of the source.

The square of the visibility magnitude is

\begin{equation}
|V_1(u,v)|^2=1+A^2+2A\cos\phi,
\end{equation}
\noindent where $\phi=2\pi(u\xi_o+v\eta_o)$.

Clearly, since $\cos\phi$ is even, the source that is symmetrical
to the original source relative to the map center will have the
same visibility magnitude. Therefore, image reconstruction from
the visibility magnitude alone is possible to within rotation
through $180^o$.

For a symmetrical source with brightness $B$ of
the symmetrical components and coordinates ($\xi_o,\eta_o$), ($-\xi_o,-\eta_o$),
the visibility magnitude is

\begin{equation}
|V_2(u,v)|=|1+2B\cos\phi|.
\end{equation}

Clearly, function (2) has singularities at zeros, and it cannot
have an analytic continuation to the entire complex plane. In this
case, the condition for the spectrum magnitude being analytic (the
existence of all derivatives)is satisfied for $B<1/2$, because the
following representation is valid:

\begin{equation}
|V_2(u,v)|=1+2B\cos\phi.
\end{equation}

Below, we are concerned only with the cases where the spectrum
magnitude of the source is an analytic function;this is achieved
only when the flux from the central component dominates over that
from the remaining components of the source. This condition is
almost always satisfied for real sources. A deviation from this
condition is possible only in the case of baseline-by-baseline
data editing.

The square of the visibility magnitude is

\begin{equation}
|V_2(u,v)|^2=1+4B\cos\phi+4B^2\cos^2\phi.
\end{equation}

The maximum and minimum values of (2)and (4) are reached at
$\cos\phi=1$ and $\cos\phi=-1$, respectively. For $A=2B$,the
minima and maxima of (2)and (4) coincide. The difference between
the squares of the visibility magnitudes for the two-component and
the corresponding three-component sources under the condition
$A=2B$ is

$$
\Delta
(V^2(u,v))=A^2-A^2\cos^2\phi=A^2\sin^2\phi.
$$

Clearly, a symmetrical object can be distinguished from an
asymmetric object only due to this difference; the higher the
relative brightness of the jet component $A$, the more reliable
the uniqueness of the solution.

Let us now complicate the source model to a three-component one
with relative brightnesses $A_1$ and $A_2$  of the compact jet
features and coordinates $(\xi_1,\eta_1)$ and $(\xi_2,\eta_2)$,
respectively. The square of the visibility magnitude for this
model can be represented as the sum

\begin{equation}
|V(u,v)|^2=1+(A_1^2+2A_1\cos\phi_1)+(A_2^2+2A_2\cos\phi_2)+2A_1A_2\cos(\phi_1-\phi_2),
\end{equation}
\noindent where $\phi_1=2\pi(u\xi_1+v\eta_1), \phi_2=2\pi(u\xi_2+v\eta_2)$.

The first term on the right-hand side of (5) represents the core,
the second term represents the first component of the jet, the
third term represents the second component of the jet, and the
fourth term represents the mutual orientation of the second and
third components that gives rise to additional components in the
autocorrelation function with coordinates equal to the difference
between the coordinates of the first and the second components of
the jet.

It is easy to show that in the general case of an $N$ -component
source,

\begin{equation}
|V(u,v)|^2=1+\sum_{i=1}^{N-1}
A_i^2+2\sum_{i=1}^{N-1}A_i\cos\phi_i+2\sum_{i=1}^{N-2}\sum_{j=i+1}^{N-1}A_iA_j\cos(\phi_i-\phi_j),
\end{equation}
\noindent where $\phi_i=2\pi(u\xi_i+v\eta_i)$.

In the case of extended features,the source is represented as a
set of point sources specified in each pixel. The visibility
function of the corresponding symmetrical (relative to the center)
source with relative brightnesses $B_i=A_i/2$ of the components is

\begin{equation}
|V(u,v)|=1+\sum_{i=1}^{N-1} 2B_i\cos\phi_i.
\end{equation}

Discarding the terms of the second order of smallness from (6) for
$A_i\ll 1$ (which is valid for most of the compact radio sources
with the core much brighter than the components of the jet), we
can roughly represent the visibility magnitude when the flux from
the central component dominates (for $\sum_{i=1}^{N-1}A_i<1$ )as

\begin{equation}
|V(u,v)|\approx 1+\sum_{i=1}^{N-1} A_i\cos\phi_i.
\end{equation}

As we see, expressions (7)and (8) are identical, to within the
discarded terms of the higher order of smallness. These terms are
important in providing unique reconstruction of the mutual
orientation of the jet components and in suppressing the symmetric
counterjet that is present on the intermediate map with a zero
spectral phase (see the next section).

As follows from expression (8), to the first approximation, each
feature of an $N$ -component source introduces a harmonic to the
visibility magnitude whose amplitude, frequency, and phase are
determined by the brightness, the distance from the center, and
the position angle of the component, respectively.

\vskip 7.5mm

\centerline{DESCRIPTION OF THE METHOD}

\vskip 7.5mm

The algorithms for image reconstruction from the spectrum
magnitude of an object, the most efficient of which are Fienup’s
algorithm and its various modifications, require knowledge of the
entire input two-dimensional sequence of sampled points. In VLBI,
however, the visibility function is measured only on a limited set
of points in the $UV$-plane, revealing large unfilled areas and a
diffraction-limited constraint. Therefore, prior reconstruction of
the visibility magnitude on the entire $UV$-plane from a limited
data set is required to successfully use existing reconstruction
algorithms like Fienup’s algorithm.

Thus,the suggested phaseless aperture synthesis method consists of
the following steps:(1)prior reconstruction of the visibility
magnitude (the object ’s spectrum)on the entire $UV$-plane, and
(2)reconstruction of the sought-for image using Fienup’s algorithm
or its modifications (Fienup 1978,1982; Bajkova 1994,1996) from
the spectrum magnitude reconstructed at the first step of the
method. The first step is performed through the reconstruction of
an intermediate image that satisfies the measured visibility
magnitude and a zero phase. Clearly, the intermediate image is
symmetric relative to the phase center of the map. The Fourier
transform of the image obtained yields the spectrum magnitude of
the source extrapolated to the region of the $UV$-plane where
there are no measurements.

Recall that Fienup’s algorithm is an iterative process of the
passage from the spatial domain of an object to the spatial
frequency domain and back (using the direct and inverse Fourier
transforms) in an effort to use the input constraints on the
spectrum magnitude of the object in the frequency domain and the
constraints on its nonnegativity and finite extent in the spatial
domain. The finite carrier within which the source is expected to
be localized should be specified in the form of a rectangle
centered at the coordinate origin, because the structure of the
source may turn out to be symmetric. Fienup et al.(1982) estimated
the size of the carrier from the autocorrelation function equal to
the inverse Fourier transform of the square of the spectrum
magnitude for the object. Very important advantages of Fienup’s
algorithm are its high stability against noise (Sanz and Huang
1983) (stability means that a small change in input data causes
the solution to change only slightly) and high speed (due to the
application of fast Fourier transform algorithms) compared to
other algorithms. The various modifications of Fienup’s algorithm
aimed at increasing the reliability of its convergence were also
developed by Bajkova (1994,1996). These papers are devoted to the
synthesis of the initial approximation that increases the
reliability of the algorithm convergence and to the prevention of
the algorithm stagnation due to computational effects.

The intermediate image can be reconstructed from the visibility
magnitude by the standard method of analytic continuation of the
spectrum using the non-linear CLEAN deconvolution procedures or
the maximum entropy method (MEM)(Cornwell et al.1999). Therefore,
an important requirement for the visibility magnitude is its
analyticity. As we noted in the previous section, this condition
is satisfied for most of the compact extragalactic radio sources
if the flux from the central compact component dominates over the
flux from the remaining fainter components.

Note an important point concerning the use of the maximum entropy
method. In contrast to the CLEAN method:the solution based on the
standard MEM is strictly positive, while the sought-for
intermediate image with a zero spectral phase is generally
alternating, taking on both positive and negative values; in
addition, it is characterized by a wider effective carrier. For
clarity, let us show this using the simple two-component model as
an example.

Let us represent the visibility magnitude (1) for a two-component
source as the following expansion into a series:

$$
|V_1(u,v)|=\sqrt{1+2A\cos\phi+A^2}=\sqrt{(1+A\cos\phi)^2+A^2\sin^2\phi}=
$$

$$
=(1+A\cos\phi)\sqrt{1+x^2}=(1+A\cos\phi)(1+\frac{x^2}{2}-\frac{1}{2}\cdot\frac{1}{4}x^4+\frac{1}{2}\cdot\frac{1}{4}\cdot\frac{3}{6}x^6-\frac{1}{2}\cdot\frac{1}{4}\cdot\frac{3}{6}\cdot\frac{5}{8}x^8+...),
$$

\noindent where $x^2=A^2\sin^2\phi/(1+A\cos\phi)^2<1$.
\vskip 2mm

Disregarding the terms higher than the fourth order of smallness
and expanding $1/(1+A\cos\phi)$ into a Taylor series, we obtain
the following representation in the form of an infinite
trigonometric series with rapidly decreasing terms:

$$
|V_1(u,v)|\approx(1+A\cos\phi)+\frac{A^2}{2}\sin^2\phi(1-A\cos\phi+A^2\cos^2\phi-A^3\cos^3\phi+...))=
$$

$$
=(1+\frac{A^2}{4}+\frac{A^4}{16})~+~(A-\frac{A^3}{8}-\frac{A^5}{16})\cos\phi~-~\frac{A^2}{4}\cos2\phi+
$$

\begin{equation}
+(\frac{A^3}{8}+\frac{A^5}{32})\cos3\phi~-~\frac{A^4}{16}\cos4\phi~+~\frac{A^5}{32}\cos5\phi~-... .
\end{equation}

Clearly, each coefficient of the cosine in expansion (9) with a
multiple argument $n\phi$ in the spatial domain is twice the
amplitude of the point component and the component symmetric to it
(relative to the center) with coordinates $(n\xi_i,n\eta_i)$ and
$(-n\xi_i,-n\eta_i)$, respectively (see also (3)). The
coefficients of the cosines with odd and even arguments are
positive and negative, respectively, suggesting that the image is
alternating.

It is easy to show that, in general, the intermediate image that
corresponds to the spectrum magnitude of a finite object and a
zero spectral phase is also an alternating function with a
theoretically unbounded carrier (since $n=\infty$).In this case,
the effective carrier of the intermediate image is much wider than
the carrier of the object, which must be properly taken into
account when reconstructing the intermediate image.

Thus, the positive definiteness of the intermediate image is
indicative of a zero phase and, hence, a symmetry of the object,
while the alternating property is indicative of a nonzero phase
and, accordingly, an asymmetry of the object relative to the
coordinate origin. Therefore, in general, since we have no prior
knowledge of whether the sought-for source is symmetric, it would
be improper to use the standard MEM with positive output to
reconstruct the intermediate image. A MEM modification suitable
for obtaining both positive and alternating solutions is required.
A generalized MEM (GMEM) suitable for reconstructing functions of
any form was developed by Bajkova (1992). In addition, in contrast
to the MEM, the GMEM yields an unbiased solution and is more
stable against noise in the data (Bajkova 2000).

It should be emphasized that the combination of the alternating
property and spatial unboundedness of the intermediate image, on
the one hand, and the constraints on the nonnegativity and finite
sizes of the object carrier, on the other hand, ensure that the
reconstruction of the sought-for image is unique.

Below, we present simulation results that confirm our conclusions.
Let us turn to Fig.1, which shows the following: {\it(a)} the
source model that consists of two Gaussian components (the central
component represents the core, and the second fainter and extended
component represents the jet); {\it(b)} the "dirty" image that
corresponds to incomplete $UV$-coverage whose Fourier transform is
equal to the visibility magnitude at the measured points and to
zero at the points without measurements (the visibility function
was generated with a signal-to-noise ratio of $\approx 10$);
{\it(c)} the intermediate image with a zero spectral phase
reconstructed from the measured visibility function using the
standard MEM with positive output; {\it(d)} the intermediate image
with a zero spectral phase reconstructed using the GMEM with
alternating output; {\it(e)} and {\it(f)} the images reconstructed
using Fienup’s algorithm from the previous MEM and GMEM images,
respectively. On all maps, the lower level of the contour line
corresponds to 0.2\% of the peak value.

As we see from the maps, the GMEM properly
intermediate-reconstructed the intermediate image with both
positive and negative values, which subsequently allowed us to
accurately reconstruct the structure of the source and, hence, its
spectral phase (to within rotation through $180^o$) using Fienup’s
algorithm. In contrast, the standard MEM algorithm with positive
output yielded no desirable result: at the output of Fienup’s
algorithm, the structure of the source was found to be nearly
symmetric (the spectral phase was virtually unreconstructed),
which is in conflict with the original model {\it(a)}. Thus, at
the first step of our mapping method, only algorithms with
alternating output, either CLEAN or GMEM, should be used to
reconstruct the intermediate image with a zero spectral phase.

An advantage of the suggested phaseless mapping method over the
methods by Baldwin and Warner (1976,1978)is the possibility of
reconstructing sources that can be represented not only as a
finite set of point components, but also sources with extended
features, which makes the suggested method more efficient.

\vskip 7.5mm

\centerline{THE RECONSTRUCTION ACCURACY}

\vskip 7.5mm

For a given reconstruction method, the accuracy of the images
being obtained depends on the following three major factors:(1)the
structure of the source being mapped,(2)the $UV$ filling
(information on the various spatial frequencies of the source),
and (3) the accuracy of the input data.

Clearly, the simpler the structure of a source at a given
resolution (e.g.,several (2 or 3)compact components with a low
dynamic range), the less stringent the requirements for the $UV$
filling and the signal-to-noise ratio of the measurements. The
more complex the structure of a source (the presence of extended
features against the background of compact features, a large
number of components, and a high dynamic range), the more complete
the filling of the spatial frequency domain (the presence of a
low-frequency domain) and the higher the signal-to-noise ratio.
General analytical expressions for estimating the image
reconstruction accuracy are very difficult to derive by nonlinear
methods, because the nonlinearity of the methods leads to
nonlinear image distortions. Therefore, it would be appropriate to
study the influence of various factors on the reconstruction
accuracy by mathematical simulation for a large number of sources
with various degrees of complexity of their structure, various
$UV$ filllings, and various specified data accuracies.

The accuracy of reconstructing the sought-for image by Fienup’s
method depends significantly on the accuracy of reconstructing the
intermediate image or, in other words, on the spectrum magnitude
of the sought-for image over the entire spatial frequency domain.
Although Fienup’s algorithms are highly stable against noise,
significant distortions of the input data may lead to
unpredictable nonlinear image distortions (the coordinates and
amplitudes of the source’s components), because the spectrum is no
longer analytic. Distortions arise,because the spatial boundaries
specified in Fienup’s algorithm within which the solution is
sought cease to correspond to the image being reconstructed. In
these cases, the sizes of the carrier should be increased to
reduce the distortions of the coordinates of the components. Our
simulations of Fienup’s hybrid input-output algorithms when the
structure of sources is approximated by simple models show that
the minimum signal-to-noise ratio of the input sampled points of
the spectrum magnitude that does noticeably distort the
coordinates of the compact components of the source, when the
coordinates of the peak values of the components are taken as
their coordinates, is about 5. In this case, the errors in the
data were simulated as additive noise with a uniform distribution.

To achieve the required data quality at the output of the first
step of the phaseless mapping method for typical "core+jet"
structures of compact extra-galactic radio sources and the $UV$
filling obtained on VLBI arrays with a small number of elements
(the cases that are the subject of discussion in this paper), the
visibility magnitude must be measured with a signal-to-noise ratio
of no less than 10. This accuracy is needed to reconstruct not
only the positively defined, but also the negatively defined
components of the intermediate image (see (9) for a two-component
source), which are relatively small, but play a major role in
suppressing the counterjets. As a result, we reconstruct the
amplitudes of the two or three brightest components of the source
with an accuracy of no less than 10\% without any apparent
distortions of their coordinates (for the mapping results of the
source 2200+420,see below)and with satisfactory suppression of the
symmetric counterjet. This is the minimum sufficient condition for
solving the restricted problem of studying the structural
evolution of compact extragalactic radio sources using
interferometers with a small number of elements. Under the
conditions of this problem, the situation where the coordinates of
the components would be noticeably and unpredictably distorted
because of the nonlinearity of the reconstruction algorithms is
totally unacceptable. In reality, the visibility magnitude in VLBI
can also be measured with a higher signal-to-noise ratio.
Therefore, below, we consider examples of image reconstruction
where the data are generated with a signal-to-noise ratio of no
less than 10. When the real data are processed, the sampled points
of the visibility function with unacceptable noise parameters can
always be discarded during prior data editing.

\vskip 7.5mm

\centerline{PHASELESS MAPPING OF THE SOURCE 2200+420}

\centerline{USING MULTIBASELINE OBSERVATIONS}

\vskip 7.5mm

The goal of this section is to demonstrate the potentialities of
phaseless mapping using the well-known radio source 2200+420 (BL
Lacertae), which exhibits several compact bright features on
milliarcsecond angular scales, as an example. Parameters of this
source can be found in the NASA/IPAC Extra- galactic Database
(NED) [25]. This source belongs to the class of BL Lacertae
objects, being its brightest representative. It exhibits
long-period variability of both flux and structure. It has been
studied quite well. The structure of the source can be judged, for
example, by the VLBA maps [25] obtained over the period 1996--2000
from 15-GHz ($\lambda=2$ cm) observations.

We constructed five maps over the period 1996-2000 using the
observations of the International Astrometric and Geodetic
Programs (NEOS) on the global International VLBI Array at a
frequency of 8.2 GHz ($\lambda=3.5$ cm).

The mapping was performed by using two independent packages:
QUASAR VLBImager developed by the author at the Institute of
Applied Astronomy (Russian Academy of Sciences)and CalTech’s
DIFMAP.

The parameters of the observations and the synthesized maps are
given in the table. It lists the dates and frequencies of the
observations, the global VLBI stations involved in the experiment,
the number of measurements, and parameters of the maps (peak
fluxes and the parameters of the Gaussian beam with which the
solution was convolved).

The maps obtained by the phaseless method (QUASAR VLBImager)
described above and by the self-calibration method in terms of
differential mapping (DIFMAP) are shown in Fig.2. In these maps
and those shown below, the minimum level of the contour line
corresponds to 1\% of the peak value.

Let us turn to the figure. The images along the rows pertain to
different dates of observations. The first column $((a)-(e))$
gives the $UV$ fillings; the second column $((f)-(j))$ gives the
intermediate images obtained from a given visibility magnitude and
a zero spectral phase (some of them $((f),(i))$ were obtained by
using the GMEM, while others $((g),(h),(j))$ were obtained by the
CLEAN method); and the third column $((k)-(o))$ gives the
sought-for images reconstructed from the previous images by using
Fienup’s algorithm. The images obtained show the structure of the
source that is immediately adjacent to the core (within 2 to 3
mas) and allows its evolution with time to be traced. We see
individual components of the jet at position angles in the range
$(-170^o,-180^o)$ whose brightnesses and positions change from
date to date. Our maps qualitatively agree with the 2-cm VLBA maps
see [26]). The fourth column of Fig.2 $((p)-(t))$ gives the maps
obtained by the adaptive calibration method using the DIFMAP
package (the scale of these maps is half the scale of the previous
ones).

Analysis of images $((p)-(t))$ shows that it is not always
justifiable to use equations for closure phases. Thus, for
example, we see a symmetric counterjet on maps $(p)$ and $(q)$,
which, as follows from the higher-visibility quality astrophysical
VLBA maps (Fig.2), should not be there. The presence of a
counterjet is indicative of incomplete spectral phase retrieval,
which shows the spurious quasi-symmetry of the source. As a
result, the strong component of the jet was not fully
reconstructed;i t broke down into two parts, distributing the
brightness between them. Thus, here, we clearly show a case where
using unreliable phase information may prove to be more dangerous
than its retrieval from the visibility magnitude measured with a
sufficient accuracy.

A comparison (given the beam size) of maps $(m)$ and $(n)$ with
maps $(r)$ and $(s)$, respectively, shows good agreement (see also
the peak fluxes in the table), implying that the visibility
function was measured reliably in these cases. A comparison of map
$(o)$ with map $(t)$ again argues for the phaseless mapping that
resolved not one, but two components of the jet present on the
maps for the two previous dates.

\bigskip
\begin{small}
\noindent{\bf Table.} Parameters of the observations and the synthesized maps for the source 2200 +420

\noindent\begin{tabular}{|c|c|c|c|c|c|}
\hline
Date &Frequency&VLBI   &Number&Package~~"QVImager"&Package~~"DIFMAP"\\
dd/mm/yy& (MHz) &stations&of UV-  &Peak flux~~~~~~~FWHM&Peak flux~~~~~FWHM\\
     &       &       &points&Jy/beam~~~~~~~~~~~~~~~~~~&Jy/beam~~~~~~~~~~~~~~~~\\
\hline
26/03/96&8210.99&$GKWN_yN_{20}F$&103&1.62~~~~~~~~$0.63\times0.63$&1.30~~~~~~$0.56\times 0.53$ \\
 & & & &                                                  &~~~~~~~~~~~~~$34.3^o$ \\
\hline
03/09/96&8182.99&$GKN_{20}WFN_y$&127&1.48~~~~~~~~$0.63\times0.63$&1.57~~~~~~$0.51\times 0.57$ \\
 & & & &                                                  &~~~~~~~~~~~~~$-1.6^o$ \\
\hline
12/11/96&8182.99&$FGN_{20}WKN_y$&100&1.79~~~~~~~~$0.63\times0.63$&1.82~~~~~~$0.55\times 0.54$ \\
 & & & &                                                  &~~~~~~~~~~~~~$40.4^o$ \\
\hline
26/08/97&8182.99&$KWN_{20}AN_y$   &83 &0.72~~~~~~~~$0.63\times0.63$&0.89~~~~~~$0.80\times 0.51$ \\
 & & & &                                                  &~~~~~~~~~~~~~$-33.0^o$ \\
\hline
06/10/98&8182.99&$KN_{20}G_gFWN_y$&115&1.31~~~~~~~~$0.63\times0.63$&1.04~~~~
~~~$0.56\times 0.43$ \\
 & & & &                                                  &~~~~~~~~~~~~~$-15.3^o$ \\
\hline
26/03/96&8210.99&$GK$&21&1.10~~~~~~~~$0.63\times0.63$& -- \\
\hline
\end{tabular}
\end{small}

\noindent {\small Abbreviated names of the stations: $A$ --
Algopark, $F$ -- Fortleza, $G$ -- Gilcreek, $G_g$ -- GGAO7108, $K$
-- Kokee, $N_{20}$ -- NRAO20, $N_y$ -- NyAlesund, $W$ --
Wettzell.}

\vskip 7.5mm

The latter result can be explained as follows. If the visibility
magnitude has been measured with a sufficiently high accuracy,
then the intermediate image of the source with a zero spectral
phase will contain all of the structural features and their mirror
features relative to the phase center of the map. The accuracy of
reconstructing the coordinates of the source’s components depends
only on the accuracy of measuring the visibility magnitude. The
subsequent image reconstruction using Fienup’s algorithm leads to
an approximately twofold enhancement of the structural components
with correct coordinates, as we see from a comparison of the maps
in the middle and the last columns of Fig.2. In contrast, using
distorted or insufficient phase information in adaptive
calibration methods may lead to an unpredictable distortion of the
source’s structure, both the brightness and the coordinates of its
individual components. In this case, the reconstruction accuracy
depends not only on the accuracy of the visibility magnitude, but
also on the degree of distortion of the phase information.

\vskip 7.5mm

\centerline{MAPPING OF COMPACT RADIO SOURCES}
\centerline{ON A TWO-ELEMENT INTERFEROMETER}

\vskip 7.5mm

During the Earth’s rotation, the interferometer baseline vector
describes an ellipse whose coordinates on the UV-plane are defined
by the equation (Thompson et al.1986)

\begin{equation}
\frac{u^2}{(L_x^2+L_y^2)}+\frac{(v-v_o)^2}{(L_x^2+L_y^2)\sin^2\delta_o}=1,
\end{equation}
\noindent where $L_x, L_y$ -- are the components of the baseline
vector $P_{ij}(u,v,w)$ along the equatorial coordinate axes,
$\delta_o$ is the declination of the phase center of the source,
and $v_o=w\cos\delta_o$. Below,we assume that the coordinates
$u,v,w$ are measured in units of wave-lengths.

In investigating a single-baseline interferometer for mapping, it
would be appropriate to use the approximation of a model source by
a set of point ponents.T his approximation is justifiable,
because, in this case, the spectrum magnitude is extrapolated to
higher frequencies from the values specified only on one curve
(10); image reconstruction in the form of a set of very compact
features in the limit of point components corresponds to this
case. Therefore, sources with this type of structure are most
suitable for observations on a two-element interferometer. If,
however, the source at a given resolution of the instrument
reveals extended features, they will be represented in the
reconstructed maps as compact components with coordinates at the
points of maximum brightness.

Clearly, the parameters $A, r, \theta$ uniquely, to within
rotation through $180^o$, determine the one-dimensional visibility
function on curve (10). Let us illustrate how variations of the
source’s parameters affect the form of the visibility magnitude
using a specific example.

In Fig.3, the left column shows the images of model sources with
the following parameters of the compact component of the jet:

\noindent$(a)$  $A=0.4$,~~$r=6.0$ mas,~~$\theta=38^o$,

\noindent$(b)$  $A=0.1$,~~$r=6.0$ mas,~~$\theta=38^o$,

\noindent$(c)$  $A=0.4$,~~$r=2.0$ mas,~~$\theta=38^o$,

\noindent$(d)$  $A=0.4$,~~$r=6.0$ mas,~~$\theta=68^o$,

\noindent$(e)$  $B=0.2$,~~$r=6.0$ mas,~~$\theta_1=38^o$,~~$\theta_2=218^o.$

All of the changes in the form of the visibility magnitude
determined on the Svetloe-Zelenchuk baseline for a source with a
declination of $\delta=73.5^o$ are shown in Fig.4. The
corresponding $UV$-coverage is shown in Fig.5$(a)$.

In all panels of Fig.4, the pluses indicate the visibility
magnitude of the reference source (time is plotted along the
horizontal axis) shown in Fig.3$a$; the crosses indicate the
visibility magnitude of the source with a changed value of a
particular parameter. Thus, we can see from Figs.$(3a--3d)$ how
changes in brightness $A$ of the component, in distance $r$ from
the map center, in position angle $\theta$, and the separation of
the component into two symmetric parts with equal brightnesses,
respectively, affect the visibility function.

More specifically, a decrease in the brightness of the component
causes a decrease in the modulation of the visibility magnitude; a
decrease in the distance of the component from the center causes a
decrease in the modulation frequency;a change in the position
angle causes a linear shift in the visibility function with time;
and the separation of the jet component into two symmetric parts
with equal brightnesses causes an increase in the visibility
magnitude by approximately $A^2\sin^2\phi/2(1+A\cos\phi)^2$.

Let us now present the results of our reconstruction of the
two-component sources shown in Fig.3 from visibility magnitude
data generated with a signal-to-noise ratio of $\approx 10$ on the
Svetloe-Zelenchuk baseline.The right column of Fig.3 shows the
final images reconstructed by using the GMEM and Fienup’s
algorithm. Analysis of our results indicates that the brightnesses
of the components were reconstructed with an accuracy up to 10\%,
and the coordinates of the components were reconstructed without
apparent distortions.

Thus, for two-component sources, mapping using single-baseline
data yields a reconstruction quality that may well be acceptable
for solving a number of problems. One of these problems might be,
for example, investigation of the motion of the brightest
components of a source on long time scales; another problem can be
allowance for the influence of the structure in astrometric
reduction, because the components that are brightest and farthest
from the core introduce the largest error in determining the
coordinates of reference sources (Bajkova 2002a).

Clearly, the more complete and accurate the determination of the
visibility magnitude on an interferometer baseline, the higher the
accuracy of reconstructing the structure of a source by using the
methods of the analytic continuation of the spectrum. In practice,
the reconstruction problem is solved with a limited accuracy,
because the data are discrete and contain measurement errors,
causing the necessary analyticity condition to be violated.
Fortunately, however, as we showed above, acceptable image
estimates can be obtained by using reconstruction methods (GMEM,
Fienup’s algorithm) that are stable against noise. In this case,
the brightest components are reconstructed reliably. Thus, data
with a signal-to-noise ratio of $\approx 10$ yield maps that
generally have two or three very bright, well-reconstructed
components.

Below, we present the results of yet another simulation (Fig.5)
that confirm our conclusions. Figure $5a$ shows the diurnal $UV$-
coverage that corresponds to the Svetloe-Zelenchuk interferometer
and the declination of the source 0212+735. Figure 5$(b)$ shows
the four-component model of 0212+735 that was roughly estimated
from existing VLBA maps (see footnote 2). Figure 5$(c)$ shows the
intermediate GMEM map reconstructed from the visibility magnitude
that was measured with a signal-to-noise ratio of $\approx 10$.
Figure 5$(d)$ presents the final reconstruction result obtained by
Fienup’s method. A comparison of the original model (Fig.5$(b)$)
and the reconstructed map (Fig.5$(d)$) indicates that we have been
able to reconstruct three of the four components;the resolution of
the component near the core was not quite as good as that on the
original map. The fourth (faintest and farthest) component was not
reconstructed because of the visibility errors in which the
contribution from this component was lost (in fact, the third and
fourth components merged together). Clearly, as the accuracy of
measuring the visibility magnitude increases, the accuracy of
reconstructing individual components can also increase. However,
as follows from our results,we can obtain maps of acceptable
quality for solving the limited range of problems outlined above
even at a relatively low signal-to-noise ratio.

\vskip 7.5mm

\centerline{PHASELESS MAPPING OF THE SOURCE 2200+420}
\centerline{USING SINGLE-BASELINE OBSERVATIONS}

\vskip 7.5mm

Let us now present the results of our mapping of the source
2200+420 using data obtained only on one interferometer baseline
(Fig.6). To this end, we separated out the observations on the
Gilcreek-Kokee baseline from the geodetic and astrometric on March
26,1996. The corresponding $UV$-coverage is shown in Fig.6$(a)$.
Parameters of the observations and the output map are given in the
last row of the table.

The image of the source reconstructed using the GMEM and Fienup’s
algorithm is shown in Fig.6$(b)$. The extent to which the input
and output data agree can be judged from Fig.6$(c)$.In this
figure, the pluses indicate the measured visibility magnitude
(time is plotted along the horizontal axis), and the crosses
indicate the visibility magnitude that corresponds to the
reconstructed image. As we see from Fig.6$(b)$, we were able to
reconstruct only the most prominent two-component structure of the
source, in close agreement with the reasoning given in the
previous section.

The derived relative fluxes from the components are close to the
fluxes from the components of image $(k)$ in Fig.2. Thus, we
conclude that mapping using single-baseline data may prove to be
quite acceptable in solving the restricted problem of studying the
structural evolution of sources that consist of several bright
compact features.

\vskip 7.5mm

\centerline{CONCLUSIONS}

\vskip 7.5mm

Despite the existence of efficient adaptive calibration methods
that use closure phases directly or indirectly (Cornwell and
Fomalont 1999) and the existence of powerful processing packages
that perform them (AIPS,DIFMAP,ASL [27]), it seems to be of
considerable interest to develop and use alternative phaseless
mapping methods for the following reasons:

(1)The spectral phase is such an important characteristic of the
image that using erroneous or insufficient phase information (at a
small number of baselines)is more dangerous than restoring it from
the spectrum magnitude measured with sufficient accuracy. Such
situations are not uncommon for VLBI observations that are not
directly intended for astrophysical mapping, but provide
invaluable data for studying the structural evolution of sources
on long time scales.

(2)The phase can be uniquely retrieved in principle from the
spectrum magnitude for multidimensional ($\ge 2$) images with a
finite carrier.

(3)Since the spectra of finite functions are analytic, the entire
function can be reconstructed from the known part of it, which is
important for VLBI.

(4)There are reliable numerical methods and reconstruction
algorithms that are stable against noise in the data.

(5)The relative simplicity of the structure of compact
extragalactic radio sources on milliarcsecond angular scales
ensures that our phaseless mapping method converges reliably and
rapidly.

(6)Available Russian-made instruments with a small number of
elements (in the limit, two-element ones) can be used to solve a
limited range of problems (e.g., to study the evolution of the
brightest components of extragalactic radio sources).

However, it is important to emphasize that we did not set the goal
of contrasting the suggested phaseless method with the traditional
VLBI mapping methods that use partial phase information (equations
for closure phases). The methods that use any properly measured
phase information are always better than the phaseless methods. In
this paper, we suggest using the phaseless methods as alternatives
to the existing methods that use phase information only in the
following two cases: (1)when the equations for closure phases are
insufficient or unavailable (a two-element interferometer) and (2)
the available phase information is unreliable, and the visibility
magnitudes were measured with a sufficient accuracy.I n
particular, the suggested method can be used for mapping based on
data from an intensity interferometer.

Thus, we have considered an efficient phaseless mapping method
that was tested both on models and on real VLBI observations. We
have presented the results of our study of a two-element
interferometer, which are of current interest even now in
connection with the putting into operation of a Russian-made
instrument based on Quasar RT-32 radio telescopes (Bajkova 2002b).

\bigskip

\centerline{REFERENCES}

\medskip

1.A.T.Bajkova, Astron. Astrophys. Trans.{\bf 1},313 (1992).

2.A.T.Bajkova, Soobshch. Inst. Prikl. Astron. Ross. Akad.
Nauk,No.62 (1994).

3.A.T.Bajkova,Izv. Vyssh. Uchebn. Zaved., Radiofiz. {\bf 39},472
(1996).

4.A.T.Bajkova,T.B.Pyatunina,and A.M.Finkel’steinn, Trudy Inst.
Prikl. Astron. Ross. Akad. Nauk, Astrometry and Geodynamics (Inst.
Prikl. Astron. Ross. Akad. Nauk, St.-Petersburg, 1997), Vol.1,
p.22 [in Russian ].

5.A.T.Bajkova, Izv.  Vyssh. Uchebn. Zaved., Radiofiz. {\bf 43},
895 (2000).

6.A.T.Bajkova, Izv. Vyssh. Uchebn. Zaved., Radiofiz. {\bf 45},187
(2002a).

7.A.T.Bajkova, Russian Conference in Memory of A.A.Pistol'kors:
Radio Telescopes RT-2002 (Pushchinskaya Radioastron. Obs. Astron.
osm. Tsentr Fiz. Inst. Akad. Nauk, Pushchino, 2002b), p.20.

8.J.E.Baldwin and P.J.Warner, Mon. Not. R. Astron. Soc. {\bf 175},
345 (1976).

9.J.E.Baldwin and P.J.Warner, Mon. Not. R. Astron. Soc. {\bf 182},
411 (1978).

10.Yu.M.BruckandL.G.Sodin, Optics Comm. {\bf 30}, 304 (1979).

11.T.J.Cornwell,R.Braun,and D.S.Briggs, Synthesis Imaging in Radio
Astronomy II, Ed.byG.B.Taylor, C.L.Carilli,and R.A.Perley, ASP
Conf.Ser. {\bf 180}, 151 (1999).

12.T.J.Cornwell and E.B.Fomalont,Synthesis Imaging in Radio
Astronomy II,Ed.byG.B.Taylor, C.L.Carilli,and R.A.Perley,ASP
Conf.Ser. {\bf 180}, 187 (1999).

13.A.M.Finkelstein and A.T.Bajkova,Preprint No.15, IPA AN SSSR
(Institute of Applied Astronomy, Academy of Sciences of the USSR,
Leningrad, 1990).

14.J.R.Fienup, Opt.Lett. {\bf 3}, 27 (1978).

15.J.R.Fienup,Appl.Opt. {\bf 21}, 2758 (1982).

16.J.R.Fienup,T.R.Crimmins,and W.Holsztynski,J. Opt.Soc.Am. {\bf
72},610 (1982).

17.R.Gerchberg and W.O.Saxton, Optik {\bf 35}, 237 (1972).

18.M.H.Hayes,IEEE Trans.Acoust.,Speech,Signal Process. {\bf 30},
140 (1982).

19.Ya.I.Khurgin and V.P.Yakovlev, Finite Functions in Physics and
Engineering (Nauka, Moscow,1971) [in Russian ].

20.A.V.Oppenheim and J.S.Lim, Proc.IEEE {\bf 69}, 529 (1981).

21.T.B.Pyatunina, A.M.Finkelstein, I.F.Surkis, et al., Trudy Inst.
Prikl. Astron. Ross. Akad. Nauk, Astrometry and Geodynamics (Inst.
Prikl. Astron. Ross. Akad. Nauk, St.-Petersburg, 1998), Vol.3,
p.259.

22.J.L.C.Sanz and T.S.Huang, J.Opt.Soc.Am. {\bf 73}, 1442 (1983).

23.Image Recovery. Theory and Application, Ed.by H.Stark
(Academic,Orlando,1987; Mir,Moscow, 1992).

24.A.R.Thompson, J.M.Moran, and G.W.Swenson, Jr., Interferometry
and Synthesis in Radio Astronomy (Wiley, NewYork, 1986;
Mir,Moscow,1989).

25.http://wwwospg.pg.infn.it/PGblazar/nedBLLac.htm.

26.http://nedwww.ipac.caltech.edu.

27.http://platon.asc.rssi.ru/dpd/asl/asl.html.

\vskip 2cm

Translated by V.Astakhov

\newpage
\pagestyle{empty}

\begin{figure}

\centerline{ \psfig{figure=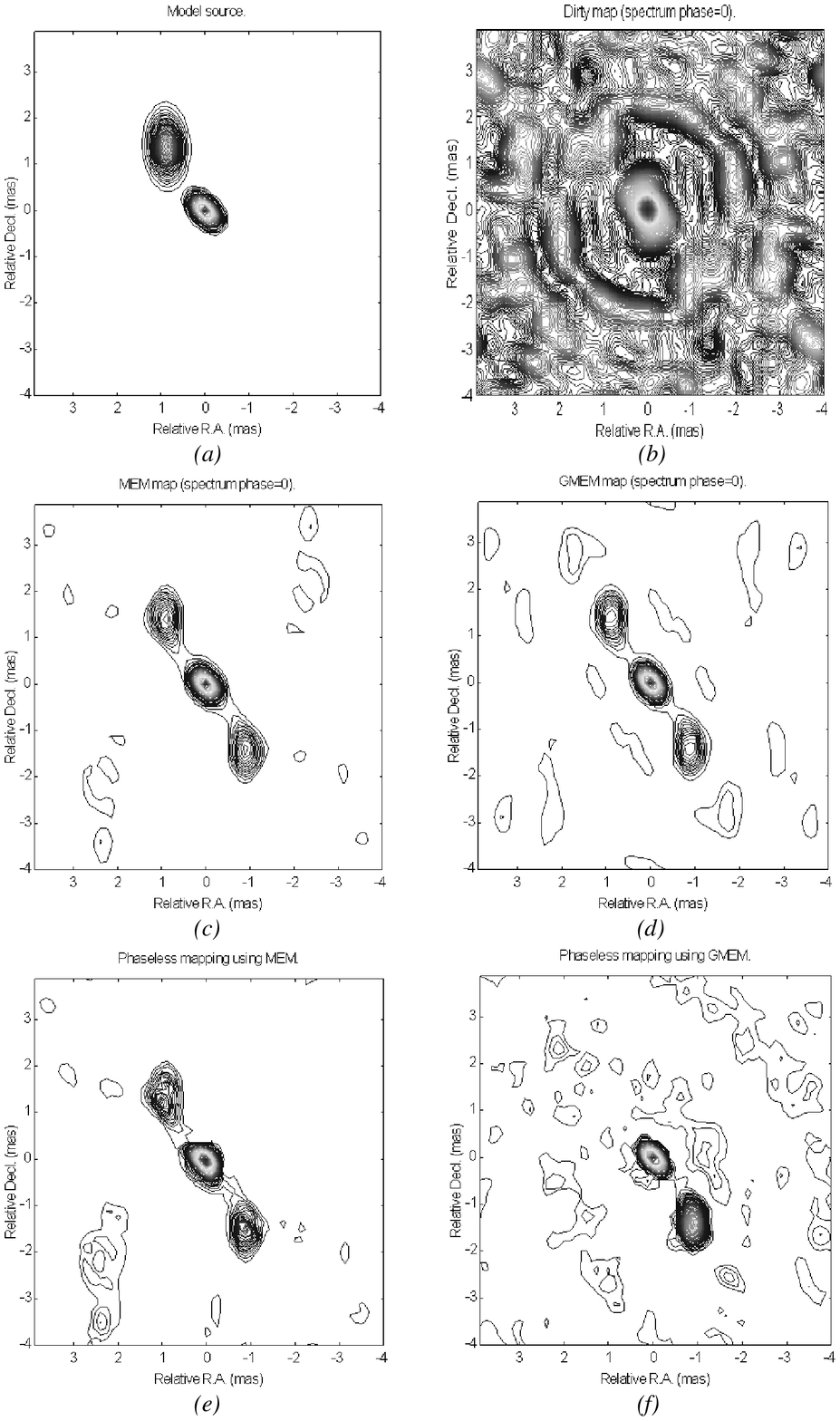,width=100mm} }

\noindent{\small Fig.1.Simulation of phaseless image
reconstruction. Comparison of the reconstruction results when
using the MEM and the GMEM (see the text)}.

\end{figure}

\newpage
\pagestyle{empty}

\begin{figure}

\centerline{ \psfig{figure=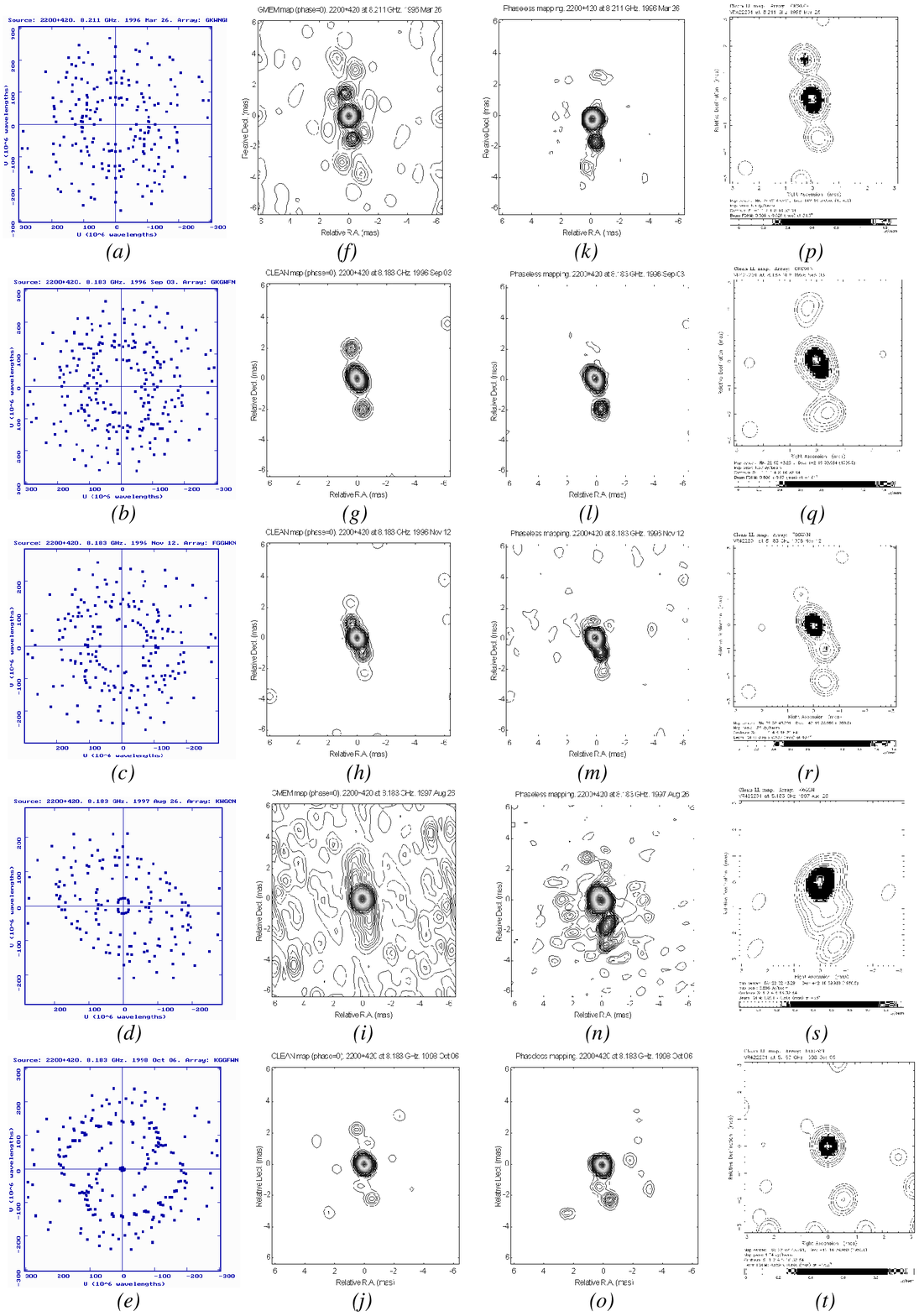,width=150mm} }

\noindent{\small Fig.2.Mapping of the source 2200 +420 based on
astrometric and geodetic data from a global VLBI array using the
phaseless (QUASAR VLBImager) self-calibration (DIFMAP) methods
(see the text). }
\end{figure}

\newpage
\pagestyle{empty}

\begin{figure}

\centerline{ \psfig{figure=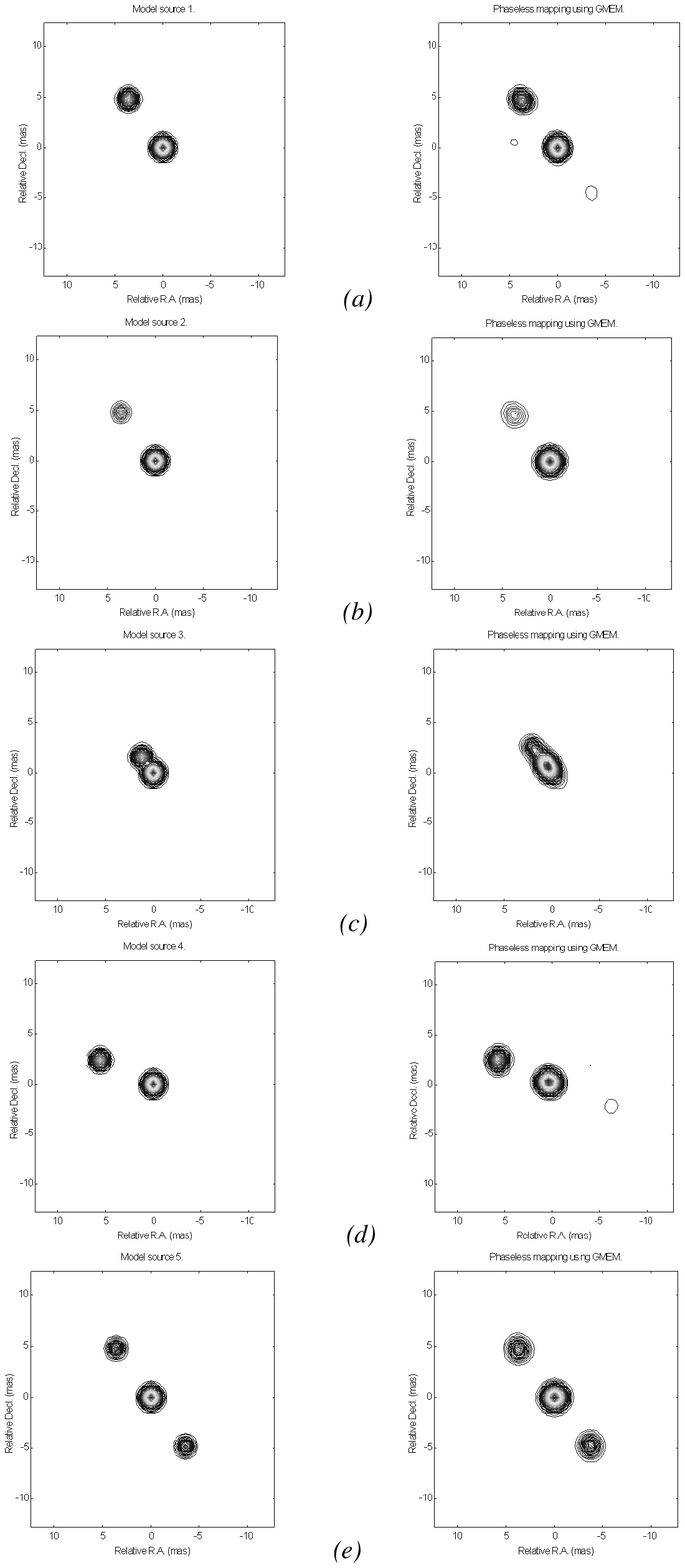,width=100mm} }

\noindent{\small Fig.3.Simulation of phaseless mapping for a
two-component source with various parameters on a two-element
interferometer (see the text).}
\end{figure}

\newpage
\pagestyle{empty}

\begin{figure}

\centerline{ \psfig{figure=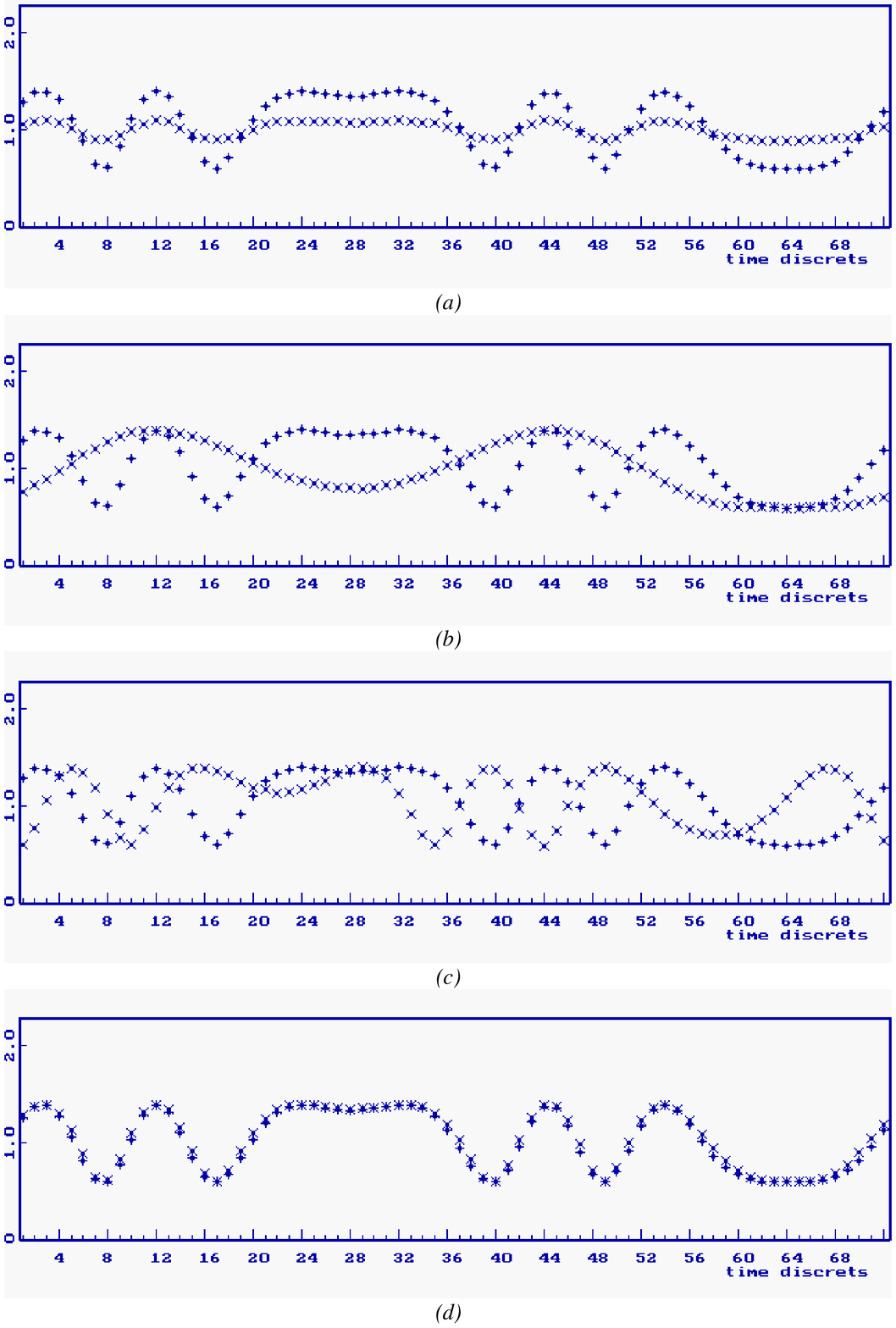,width=150mm} }

\noindent{\small Fig.4.Influence of the parameters of a
two-component source on the form of the visibility magnitude on an
interferometer baseline (see the text). }

\end{figure}

\newpage
\pagestyle{empty}

\begin{figure}

\centerline{ \psfig{figure=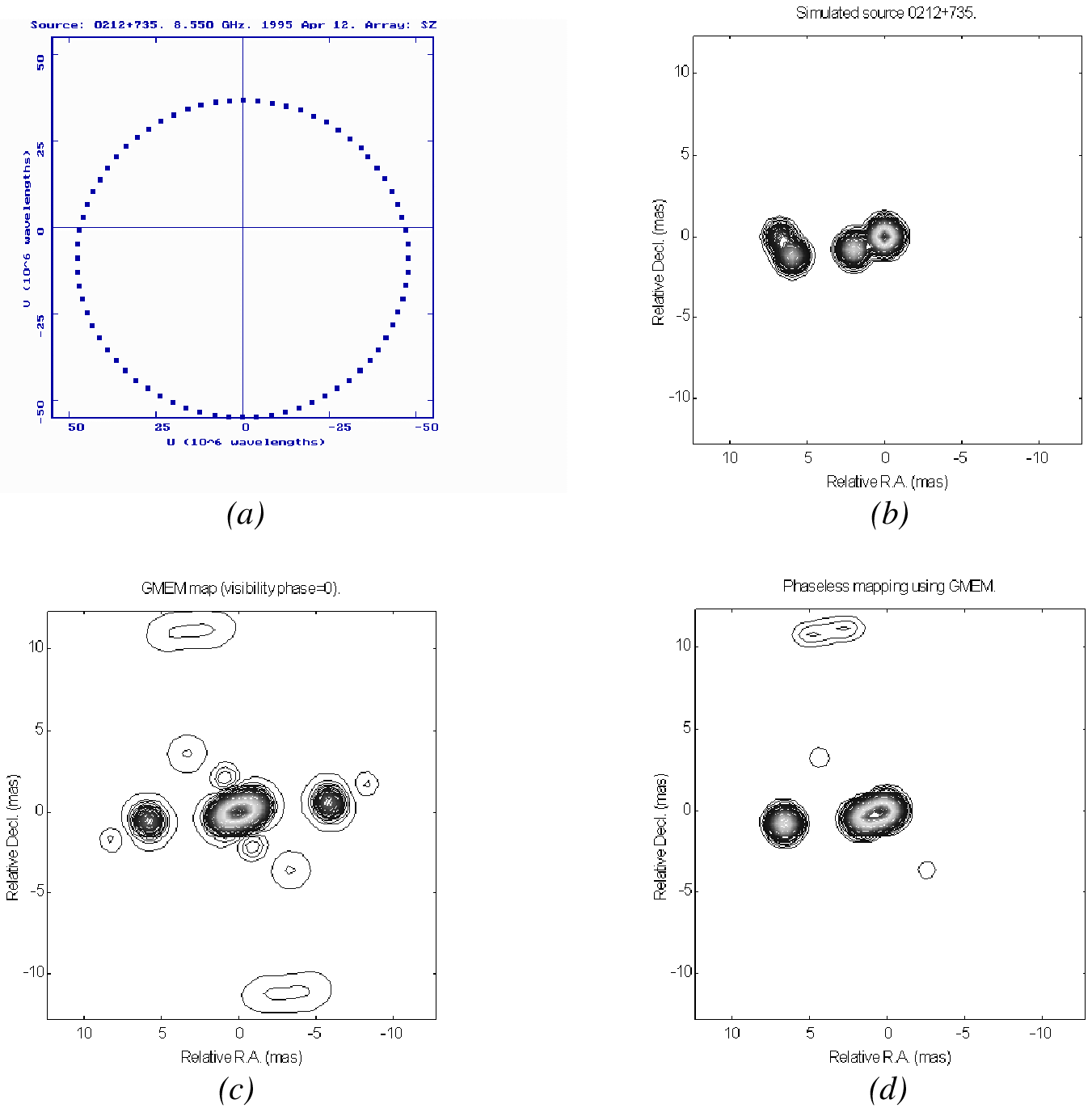,width=150mm} }

\noindent{\small Fig.5.Mapping simulation for the four-component
source 0212+735 based on observations on the Svetloe-Zelenchuk
interferometer (see the text).}

\end{figure}

\newpage
\pagestyle{empty}

\begin{figure}

\centerline{ \psfig{figure=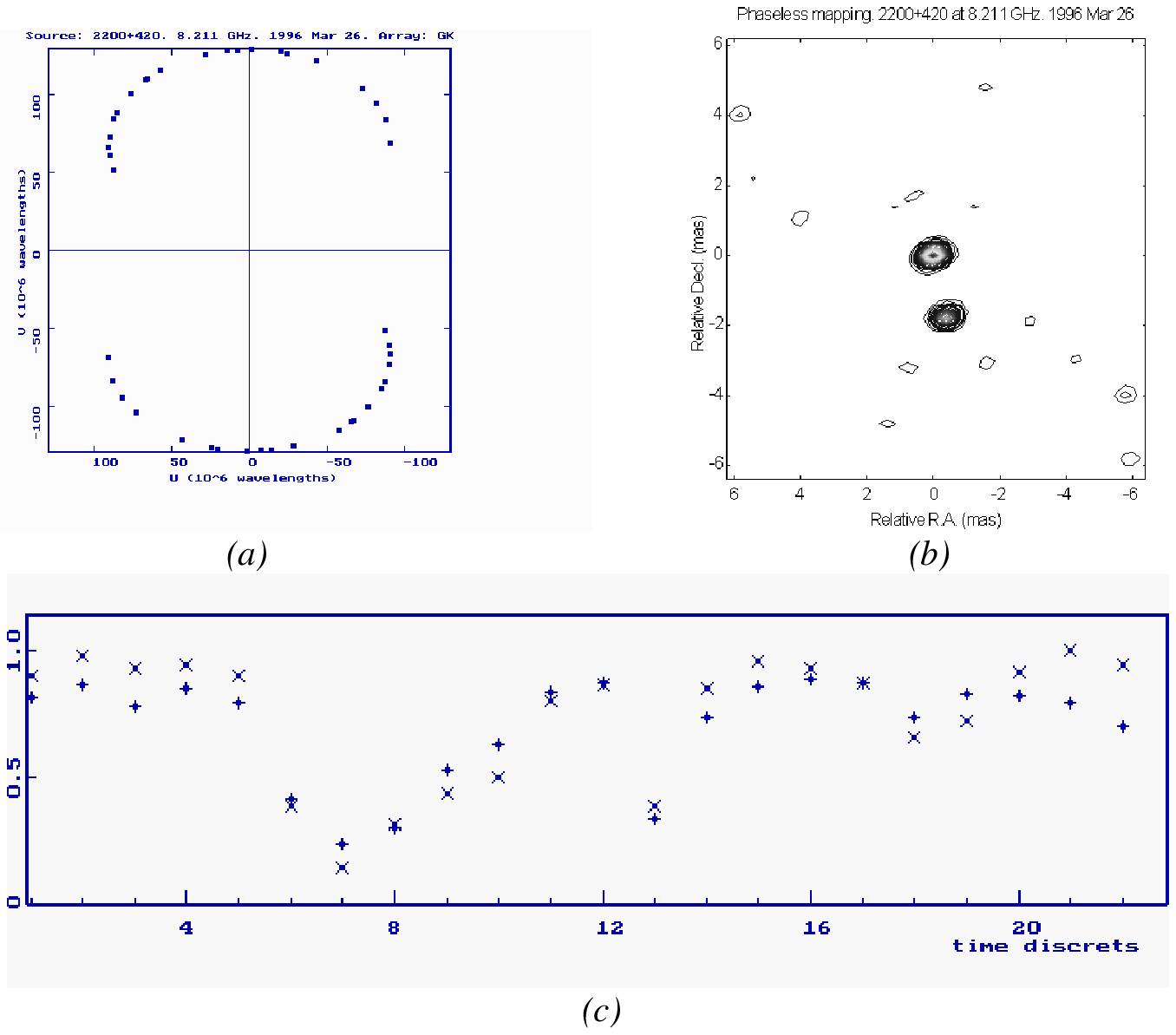,width=150 mm} }

\noindent{\small Fig.6.Phaseless mapping of the source 2200+420
using the astrometric and geodetic VLBI observations of March
26,1996, on the Gilcreek-Kokee baseline (see the text). }

\end{figure}

\end{document}